\documentclass[12pt]{iopart}
\usepackage{iopams}
\begin{document}

\title[]{Travelling waves in the expanding spatially homogeneous space-times}

\author{George Alekseev}

\address{Steklov Mathematical Institute of Russian Academy of Sciences,
\\[-0.5ex]
{\small\emph{Gubkina 8, Moscow 119991, Moscow, Russia,}}}
\ead{G.A.Alekseev@mi.ras.ru}
\vspace{10pt}
%\begin{indented}
%\item[]August 2014
%\end{indented}

\begin{abstract}
Some classes of the so called ``travelling wave'' solutions of Einstein and Einstein - Maxwell equations in General Relativity and of dynamical equations for massless bosonic fields in string gravity in four and higher dimensions are presented. Similarly to the well known pp-waves, these travelling wave solutions may depend on  arbitrary functions of a null coordinate which determine the arbitrary profiles and polarizations of the waves. However, in contrast with pp-waves, these waves do not admit the null Killing vector fields and can exist in some curved (expanding and spatially homogeneous) background space-times, where these waves propagate in certain directions without any scattering. Mathematically, some of these classes of solutions arise as the fixed points of Kramer-Neugebauer transformations for hyperbolic integrable reductions of the mentioned above field equations, or, in the other cases, -- after imposing of the ansatz that these waves do not change the part of spatial metric transversal to the direction of wave propagation. It is worth to note that strikingly simple forms of all presented solutions make possible a consideration of nonlinear interaction of these waves with the background curvature and singularities as well as a collision of sandwiches of such waves with solitons or with each others in the backgrounds where such travelling waves may exist.
\end{abstract}

%\maketitle

\section*{Introduction}
Among many exact solutions which were discovered by different authors during almost a hundred years' history of General Relativity (the most part of these can be found in the books \cite{SKMHH:2003}--\cite{Griffiths-Podolsky:2009}), there are very few classes of explicitly known solutions of Einstein's field equations in General Relativity and in other gravity models which can shed some light on various nonlinear wave phenomena which characterize strong gravitational, electromagnetic or other types of waves propagating through (and interacting with) some non-trivial background space-time geometries and external fields.

From physical point of view, consideration of any nonlinear wave phenomena becomes more clear if we can distinguish the waves which take part in the interactions. This is, for example, in the case of collision of waves with distinct wavefronts in which certain kind of waves approach each other before their collision, or in the case of interaction of incident waves of infinite duration but rapidly decreasing amplitudes in which the parameters of initial waves can be determined only asymptotically, as well as in the case of decay of some initial field configuration producing some outgoing waves. In such cases, we can say about some rather simple kinds of waves which we call here as ``travelling waves'' and which propagate in some space-time region in certain direction without scattering and without creation of focussing singularities.

In the literature of the last decades a lot of attention was concentrated on the studies of one type of such travelling waves -- plane gravitational and electromagnetic waves which represent a subclass of the class of plane-fronted waves with parallel rays or ``pp-waves''. These solutions (discovered long ago for vacuum by Brinkman \cite{Brinkman:1925} and then considered for Einstein-Maxwell case by Baldwin and Jeffery \cite{Baldwin-Jeffery:1926}) are algebraically special and the corresponding space-time geometries admit shear-free, twist free and expansion-free null congruences and possess rather large groups of isometries including the null Killing vector field. Apart from other pp-waves, the plane wave solutions are characterized by a supplemental condition that for these waves various physical and geometrical parameters are constant along the wavefronts. Later, the plane-wave solutions were presented by Rosen \cite{Rosen:1937} in a very convenient general form which show explicitly the transversal character of these waves. In this form all components of the solutions depend only on a null coordinate which labels the wavefronts. It is remarkable that in General Relativity the class of plane wave solutions for gravitational waves in vacuum as well as for gravitational and electromagnetic waves in electrovacuum case depend on arbitrary functions of a null coordinate which determine the amplitudes and polarizations of waves. Despite  very simple explicit form of plane-wave solutions, these may have rather interesting and non-trivial geometrical structures (see \cite{SKMHH:2003}, \cite{Griffiths-Podolsky:2009} for many details and references).

Because of the existence of a null Killing vector field and other symmetries of plane waves, these solutions can be matched easily with the Minkowski background on the wavefronts. This leads  to a construction of more realistic solutions for wave pulses which propagate through the Minkowski background with distinct wavefronts and, in particular, for sandwich waves. However, for a description of waves which propagate in the Minkowski background and which possess some non-planar wavefront geometry, or of the waves propagating in some non-trivial curved backgrounds in which the curvature characteristics inevitably evolve along the wavefronts, we need to consider the travelling waves of the other types.

Many particular solutions for travelling waves (different from plane waves) are known within the class of vacuum space-times which metrics are diagonal and depend on time and one spatial coordinate. These metrics describe propagation of gravitational waves with constant linear polarization. As it is well known, the Einstein equations for these metrics reduce to a linear equation for one real unknown function -- the Euler-Poisson-Darboux equation which admits many explicit solutions for travelling waves. In particular, the examples of solutions for such waves with distinct wavefronts and different level of smoothness of metric on these fronts, propagating in different cosmological backgrounds can been found in \cite{Alekseev-Griffiths:1995},
the waves propagating through the Minkowski background with different forms of the wavefronts (plane, cylindrical, spherical, elliptical, etc) were considered in \cite{Alekseev-Griffiths:1996, Alekseev-Griffiths:1997}.
More general solutions for such ``linear'' travelling gravitational waves with constant linear polarization propagating in some nontrivial backgrounds can arise as some specific solutions of the corresponding characteristic initial value problem formulated for the Euler-Poisson-Darboux equation. The general solution of the characteristic initial problem for this equation  was described in terms of the Abel transform in \cite{Hauser-Ernst:1989a}, \cite{Hauser-Ernst:1989b} and later was discussed in \cite{Griffiths:1991}, \cite{Griffiths-Santano-Roco:2002}.

Much less solutions are known for travelling waves which do not possess a constant linear polarization and/or include, besides gravitational part, also the electromagnetic or some other field components. Most of the known examples of such solutions belong to the classes of fields which components depend on two coordinates only -- the time and one spatial coordinate or, more precisely, to the classes of space-times which metric admits a two-dimensional (in 4D space-times) or, in general, (D-2)-dimensional (in D-dimensional space-time) Abelian isometry group with space-like orbits, and all matter fields and their potentials share this symmetry.
With this space-time symmetry ansatz the Einstein's field equations remain nonlinear, but in many physically important cases these symmetry reduced equations (similarly to elliptical Ernst equations for stationary axisymmetric fields  \cite{Ernst:1968a,Ernst:1968b}) can be expressed conveniently in the form of the hyperbolic Ernst equations (the Einstein equations for vacuum, electrovacuum Einstein - Maxwell equations and some others) or some matrix analogues of the Ernst equations (the equations for bosonic dynamics of some string gravity models in four and higher dimensions). In these cases, the field equations occur to be completely integrable. The integrability of these equations allows to use for construction of their solutions  various symmetry transformations, soliton generating techniques, the monodromy transform approach and various integral equation methods (Concerning these methods, see the survey \cite{Alekseev:2011} for more details and references.)
These methods provide us with general schemes for solving the mentioned above integrable reductions of Einstein's field equations, principle algorithms for solution of various initial, characteristic and boundary value problems for these equations and practical methods for construction of special kinds of particular solutions (such as solitons and some others). However, the explicit realizations of these schemes and algorithms is very difficult for more or less generic ``input data'' for the solutions and these can be realized only for special very particular choices of the data. This leads to explicit construction of families of solutions which include only finite number of new parameters, but not any arbitrary functions.

On the other hand, some solutions for travelling waves which are different from plane waves, propagating in curved space-times and depending on arbitrary functions of a null coordinate are known. First of all we have to mention here a class of  vacuum and stiff matter fluid solutions for gravitational wave pulses which was found and investigated in detail in \cite{Wainwright:1979}. The solutions of this class describe the waves which have no the linear polarization and propagate in a spatially homogeneous (Kasner) universe of special form. These solutions depend explicitely on one arbitrary function of a null coordinate which determine the profiles of the waves.

Another known example of solutions for travelling waves (different from plane waves) is the class of solutions for cylindrical pure electromagnetic waves which was found in \cite{Misra-Radhakrishna:1962} and mentioned also in the eq. (22.59) in the ``Cylindrical Waves'' chapter of \cite{SKMHH:2003}.
These solutions depend on two arbitrary functions of a null coordinate which determine arbitrary amplitudes of each of two states of polarization. It is necessary to note here, however, that the cylindrical wave interpretation of these solutions is not completely appropriate because of singular behaviour of these metrics on the axis $\rho=0$.

The purpose of this paper is a construction of solutions for travelling waves (different from pp-waves) depending on some arbitrary functions of a null coordinate which describe in General Relativity and in some string gravity models in four and higher dimensions propagation in some curved space-times of gravitational, electromagnetic and other massless bosonic fields with arbitrary  profiles and polarizations. For construction of solutions for such waves, we suggest two methods based on two different ansatzes which can be used for searching the travelling wave solutions as the solutions of the hyperbolic Ernst equations in General Relativity and the matrix analogies of the Ernst equations which arise in some (symmetry reduced) string gravity models.

One way in which the travelling wave solutions may arise, is a consideration of fixed points of the (generalized) Kramer-Neugebauer transformation of the solution spaces of hyperbolic integrable reductions of Einstein's field equations.
This transformation of vacuum solutions to vacuum solutions for stationary axisymmetric fields was discovered by Kramer and Neugebauer -- see \cite{SKMHH:2003} for more details and references. It is easy to see that similar transformations exist in the space of 4D vacuum solutions admitting two-dimensional Abelian isometry group with 2-surface orthogonal space-like orbits (or, equivalently, in the space of solutions of hyperbolic vacuum Ernst equation). It is surprising, may be, but rather simple calculation of the fixed points of this transformation lead to the class of solutions which depend on an arbitrary function of null coordinate and which admit unambiguous interpretation as travelling waves propagating on some special case of Kasner background. And, this class of solutions coincides with the class of Wainwright vacuum metrics mentioned above. Thus, we do not obtain new solutions for this simplest case, but this suggests the way for finding similar solutions for travelling waves on some spatially homogeneous backgrounds for other integrable hyperbolic symmetry reductions of Einstein's field equations. This was confirmed below by a construction of traveling wave solutions for hyperbolic symmetry reductions of dynamical equations for massless bosonic fields in string gravity.

Another ansatz also leads to construction of some classes of travelling wave solutions for not gravitational, but pure electromagnetic or other massless bosonic fields in four and higher dimensions which depend on a set of arbitrary functions of a null coordinate. This ansatz restricts the consideration by the solutions for plane-fronted waves which propagate on the expending spatially homogeneous background and for which the part of spatial metric transversal to the direction of waves propagation remains unperturbed. In the case of Einstein-Maxwell fields in four dimensions this ansatz leads unambiguously to the  plane-fronted pure electromagnetic travelling waves propagating in the symmetric Kasner space-time along its axis of symmetry.
The solutions of this class depend on two arbitrary functions of a null coordinate which determine the arbitrary amplitudes, profiles and polarizations of these waves. We note here that formally the class of waves derived in this way is a ``twin'' of the mentioned above class of solutions for cylindrical electromagnetic waves propagating on some static background found in \cite{Misra-Radhakrishna:1962}, but its interpretation as cylindrical waves is not appropriate because of singular behaviour of solutions on the axis of symmetry. In our case, we have the plane-fronted waves propagating on the non-static,  expandig (Kasner) background and the existence of Kasner initial singularity in these solutions is justified from physical point of view.

In this paper we use also the ansatz similar to the mentioned just above for construction of travelling waves solutions for some massless bosonic fields in string gravity which also propagate in the expending spatially homogeneous backgrounds in four and higher dimensions and possess arbitrary profiles and polarizations.

In the next section of this paper we recall at first the hyperbolic vacuum Ernst equation and the corresponding real hyperbolic form of the Kramer-Neugebauer transformations of its solution spaces. Then we describe the derivation of solutions which arise as the fixed points of this transformation. After that, in another section we recall the hyperbolic electrovacuum Ernst equations in the Einstein-Maxwell theory in four dimensions and apply to these equations the second ansatz. Then we describe a derivation of the corresponding class of pure electromagnetic travelling waves propagating in the expending Kasner background.

In the subsequent sections, we use the same methods for construction of solutions for massless bosonic gauge fields in string gravity in four and higher dimensions. These solutions describe the travelling waves with arbitrary profiles and polarizations propagating in some expanding spatially homogeneous backgrounds. In the Conclusions we summarize the results and outline the related interesting questions.

\section*{Hyperbolic vacuum Ernst equation and Kramer-Neugebauer transformation of its solution space}

\subsection*{The space-time symmetry ansatz}
It is well known that in any 4D space-time which admit the  two-dimensional Abelian orthogonally transitive isometry group with space-like orbits, the coordinates can be chosen in such a way that all metric components depend only on time  and one spatial coordinate, say $x^0=t$ and $x^1=x$, or, equivalently, on null cone coordinates $u=t-x$ and $v=t+x$ and metric can be presented in the form
\begin{equation}\label{Metric}
ds^2=-f\, du dv+g_{ab} dx^a dx^b,
\end{equation}
where $a,b,\ldots=2,3$ and $\{x^0,\,x^1,\,x^2,\,x^3\}=\{t,x,y,z\}$; the conformal factor $f$ and the remaining metric components $g_{ab}$ are functions of $u$ and $v$ only. It is convenient to  parameterize the metric components $g_{ab}$ by three real functions $\alpha$, $H$ and $\Omega$ so that
\begin{equation}\label{gab}
g_{ab} dx^a dx^b=H(dy+\Omega\, dz)^2+\frac{\alpha^2}{H} dz^2
\end{equation}
where $H>0$, $\alpha> 0$ and $\det\Vert g_{ab}\Vert =\alpha^2$.

\subsection*{The Ernst potential and hyperbolic Ernst equation for vacuum metrics}
Similarly to stationary axisymmetric vacuum metrics for which the Ernst potential was introduced in \cite{Ernst:1968a}, the metrics
(\ref{Metric}) can be characterized also by a complex scalar potential $\mathcal{E}$ defined up to an imaginary constant by the relations
\begin{equation}
     \label{EPotential}
\mbox{Re}\,{\cal E}=-H,\qquad d(\mbox{Im}\,{\cal E})=\alpha^{-1} H^2\,
{}^{\ast}\!d\Omega,\qquad {}^{\ast}\!d\equiv du\,\partial_u-dv\,\partial_v
\end{equation}
where ${}^{\ast}\!d$ is the Hodge star operator on the plane $(u,v)$. In accordance with vacuum Einstein equations for metrics (\ref{Metric}), the functions $\mathcal{E}(u,v)$ and $\alpha(u,v)$ should satisfy the nonlinear Ernt equation and the d'Alembert equation respectively:
\begin{equation}\label{ErnstEqn}
\left\{\hskip-1ex\begin{array}{l} \mbox{Re}\,{\cal
E}\left(2\,{\cal E}_{uv}+ \displaystyle{\frac
{\alpha_u}\alpha}\,{\cal
E}_v+\displaystyle{\frac{\alpha_v}\alpha}\,{\cal
E}_u \right)-2\, {\cal E}_u \,{\cal E}_v=0\\[1ex]
\,\,\alpha_{uv}=0.
\end{array}\right.
\end{equation}
where all suffices mean the derivatives. Given a solution $(\alpha,\,{\cal E})$ of (\ref{ErnstEqn}), the functions $H$ and $\Omega$ can be determined from  (\ref{EPotential}), while the function $f$ is determined by the relations:
\begin{equation}\label{VFactor}
\displaystyle{\frac {f_u}f}=\displaystyle{\frac{\alpha_{uu}}
{\alpha_u}}-\displaystyle{\frac {H_u}H}+
\displaystyle{\frac \alpha {2\alpha_u}}\displaystyle{\frac
{\left\vert{\cal
E}_u\right\vert^2}
{H^2}},\qquad
\displaystyle{\frac {f_v} f}=\displaystyle{\frac {\alpha_{vv}}
{\alpha_v}}-\displaystyle{\frac {H_v} H}+\displaystyle{\frac \alpha
{2\alpha_v}}\displaystyle{\frac {\left\vert{\cal
E}_v\right\vert^2}{H^2}}.
\end{equation}
For different choices of the solution for $\alpha(u,v)$, the gauge freedom $u\to h(u)$, $v\to g(v)$ remaining in (\ref{Metric}) allows to choose $\alpha$ as a new time-like or spatial coordinate $t$ or $x$ respectively. However, in some contexts, such as consideration of colliding plane gravitational waves with distinct wavefronts propagating through the Minkowski background, one may need to consider more complicate solutions for $\alpha(u,v)$ determined by the profiles of waves before their collision (see \cite{Griffiths:1991, Alekseev-Griffiths:2004} and the references there).

\subsection*{Alternative vacuum Ernst potential and  Kramer-Neugebauer transformation} As it can be checked by a direct calculation,  the following combination of vacuum metric functions (\ref{gab}) also should satisfy the hyperbolic Ernst equation (\ref{ErnstEqn}):
\begin{equation}\label{AltEPotential}
{\cal E}=-\displaystyle{\frac{\alpha}{H}}+i \Omega
\end{equation}
Therefore, given a solution $(\alpha,{\cal E})$ of (\ref{ErnstEqn}), one can construct two vacuum metrics corresponding to different ``interpretations'' (\ref{EPotential}) and (\ref{AltEPotential}) of vacuum Ernst potential:
\begin{equation}\label{AltComponents}
\fl\left\{\begin{array}{l}
H_{(1)}=-\mbox{Re}\,{\cal E}\\
d\,\Omega_{(1)}=\alpha\,(\mbox{Re}\,{\cal E})^{-2}\, {}^{\ast}\!d (\mbox{Im}\,{\cal E}),
\end{array}\right.\qquad
\left\{\begin{array}{l}
H_{(2)}=-\displaystyle\frac{\alpha}{\mbox{Re}\,{\cal E}},\\
\Omega_{(2)}=\mbox{Im}\,{\cal E},
\end{array}\right.\qquad f_{(2)}=\frac{H_{(1)}}{\sqrt{\alpha}} f_{(1)}
\end{equation}
Accordingly, one can consider the transformation acting on the space of solutions of the described above symmetry reduced vacuum Einstein equations which take the form
\begin{equation}\label{NKTransform}
\{H_{(1)},\,\Omega_{(1)}\}\quad\longleftrightarrow\quad \{H_{(2)},\,\Omega_{(2)}\}
\end{equation}
(Such transformations of vacuum solutions for stationary axisymmetric fields was discovered by Neugebauer and Kramer -- see \cite{SKMHH:2003} for more details and references).

\section*{Travelling vacuum gravitational waves which are fixed points of Kramer-Neugebauer transformation}
Consider now the question, whether the Kramer-Neugebauer transformation possess the fixed points, i.e. whether the solutions exist which are invariant under this transformation of the solution space. It is clear that these solutions should satisfy
\begin{equation}\label{Invariance}
H_{(1)}=H_{(2)},\qquad \Omega_{(1)}=\Omega_{(2)}.
\end{equation}
These conditions lead, in accordance with (\ref{AltComponents}), to the relations:
\begin{equation}\label{VacuumWaves}
\mbox{Re}\,{\cal E}=-\sqrt{\alpha},\qquad
d (\mbox{Im}\,{\cal E})={}^{\ast}\!d (\mbox{Im}\,{\cal E})
\end{equation}
The second relation here (due to the definition of ${}^{\ast}\!d$ in (\ref{EPotential})) reduces to $\partial_v (\mbox{Im}\,{\cal E})=0$ and the class of invariant solution is determined by the Ernst potential
\begin{equation}\label{VacuumWaves}
{\cal E}=-\sqrt{\alpha}+i\psi_+ (u),\qquad \alpha_{uv}=0
\end{equation}
where $\alpha$ is an arbitrary solution of the last of the above equations and $\psi_+(u)$ is an arbitrary function of $u$. The corresponding class of vacuum metrics takes the form
\begin{equation}\label{VacuumLeftWaves}
\fl \qquad ds^2=-\alpha^{-3/8}\exp\left[\displaystyle\int\limits_{u_0}^u \displaystyle\frac{{\psi_{+}^\prime\!}^2(u)}{2\alpha_u} du\right] \alpha_u  \alpha_v du dv+\sqrt{\alpha}\Bigl[dy+\psi_+(u) dz\Bigr]^2+\alpha^{3/2} dz^2
\end{equation}
Besides that, instead of (\ref{Invariance}), a slightly different invariance conditions $H_{(1)}=H_{(2)}$, $\Omega_{(1)}=-\Omega_{(2)}$ lead to
another class of vacuum metrics which possess the form
\begin{equation}\label{VacuumRightWaves}
\fl \qquad ds^2=-\alpha^{-3/8}\exp\left[\displaystyle\int\limits_{v_0}^v \displaystyle\frac{{\psi_{-}^\prime\!}^2(v)}{2\alpha_v} dv\right] \alpha_u \alpha_v du dv+\sqrt{\alpha}\Bigl[dy+\psi_-(v) dz\Bigr]^2+\alpha^{3/2} dz^2
\end{equation}
One can check easily that the metrics (\ref{VacuumLeftWaves}) and (\ref{VacuumRightWaves}) satisfy the vacuum Einstein equations.
For different choices of  $\alpha$ as a solution of the equation $\alpha_{uv}=0$ for which  $\alpha=const$ are time-like or space-like surface, the appropriate transformations $u\to h(u)$, $v\to g(v)$ allow to choose $t=\alpha$ or $x=\alpha$ respectively. The case $\alpha=t$ is the most interesting because (\ref{VacuumLeftWaves}) with $\psi_+(u)=0$ as well as (\ref{VacuumRightWaves}) with  $\psi_-(u)=0$ represent a cosmological Kasner solution with a special set of exponents $(p_x,p_y,p_z)=(-\frac{3}{13},\frac{4}{13},\frac{12}{13})$\footnote{The usual form of the family of vacuum cosmological Kasner solutions is $ds^2=-d\tau^2+\tau^{2 p_{(x)}} dx^2 +\tau^{2 p_{(y)}} dy^2+\tau^{2 p_{(z)}} dz^2 $, where the exponents should satisfy $p_{(x)}+p_{(y)}+p_{(z)}=1$, $p_{(x)}^2+p_{(y)}^2+p_{(z)}^2=1$. However, in the form (\ref{Metric}) this family of solutions is described by the expression $ds^2=t^{2 k_{(x)}}(-dt^2+dx^2)
+t^{2 k_{(y)}} dy^2+t^{2 k_{(z)}} dz^2$
where $k_{(x)}=p_{(x)}/(1-p_{(x)})$, $k_{(y)}=p_{(y)}/(1-p_{(x)})$, $k_{(z)}=p_{(z)}/(1-p_{(x)})$.} and  therefore, for $\psi_+(u)\ne 0$ or $\psi_-(v)\ne 0$ these solutions describe the plane-fronted travelling gravitational waves which profiles are determined by the arbitary functions $\psi_+(u)$ in (\ref{VacuumLeftWaves}) and $\psi_-(v)$ in (\ref{VacuumRightWaves}) and which propagate on this specific Kasner background in the positive and negative directions of the $x$-axis respectively.

In the case $\alpha=t$, the classes of solutions (\ref{VacuumLeftWaves}) and (\ref{VacuumRightWaves}) are already known. The physical and geometrical properties of these solutions as the solutions for the plane-fronted waves on a specific Kasner background were considered in detail in \cite{Wainwright:1979}. These solutions certainly can be used in farther investigation of propagation of  waves with arbitrary amplitudes in curved space-times as well as their collision and nonlinear interaction in this background.  Besides that,  it is also important for us here, that the classes of solutions (\ref{VacuumLeftWaves}) and (\ref{VacuumRightWaves}) in the above considerations arose from the ansatz of invariance of the solution under the Kramer-Neugebauer transformation of the solution space. This suggests the opportunity to use the similar ansatz for construction of classes of travelling wave solutions for integrable reductions of some other gravity models in four and higher dimensions.

\section*{Hyperbolic Ernst equations for electrovacuum fields}

\subsection*{The space-time symmetry ansatz} For 4D electrovacuum space-times which metric satisfies the same space-time symmetry ansatz as vacuum metrics discussed above and which electromagnetic potential shares the same  symmetry, the metric and electromagnetic potential can be taken in the form
\begin{equation}\label{EVMetric}
ds^2=-f\, du dv+g_{ab} dx^a dx^b,\qquad \mathbf{A}=\{0,\,0,\,A_a\}
\end{equation}
where $x^a=\{y,z\}$ and $g_{ab}$, $f$ and $A_a$ are functions of $u$ and $v$ only. We use here the null coordinates $\{u,v\}=\{t-x,t+x\}$ and the same parametrization (\ref{gab}) for $g_{ab}$ in terms of the functions $\alpha>0$, $H>0$ and $\Omega$ with $\det\Vert g_{ab}\Vert =\alpha^2$.

\subsection*{Electrovacuum Ernst potentials and hyperbolic Ernst equations} Similarly to stationary axisymmetric fields \cite{Ernst:1968b}, the electrovacuum Einstein - Maxwell fields, which satisfy the symmetry ansatz mentioned just above, admit a complete description in terms of two complex Ernst potentials ${\cal E}(u,v)$ and $\Phi(u,v)$. In the hyperbolic case, these potentials are determined by the relations:
\begin{equation}
\fl     \label{EFPotentials}
\left\{\begin{array}{rl}
\mbox{Re}\,{\cal E}=&-H-\Phi\overline{\Phi}\\
d(\mbox{Im}\,{\cal E})=&\alpha^{-1} H^2\,
{}^{\ast}\!d\Omega+ i(\overline{\Phi}d\Phi-\Phi
d\overline{\Phi})
     \end{array}\right.\hskip0.5ex
     \left\{\begin{array}{rl}
\mbox{Re}\,\Phi=&A_y\\
d(\mbox{Im}\,\Phi)=&\alpha^{-1} H(\Omega\,
{}^{\ast}\!d A_y-{}^{\ast}\!d A_z)
     \end{array}\right.
     \end{equation}
where the star operator ${}^{\ast}\!d\equiv du\,\partial_u-dv\,\partial_v$, and
$A_y$, $A_z$ are the non-vanishing components of a real electromagnetic
vector potential. In accordance with electrovacuum Einstein-Maxwell equations for the fields (\ref{EVMetric}), the functions $\mathcal{E}(u,v)$, $\Phi(u,v)$ and $\alpha(u,v)$ should satisfy the nonlinear Ernst equations and the d'Alembert equation respectively:
\begin{equation}\label{ErnstEqs}
\fl\hskip0ex\left\{\hskip-1ex\begin{array}{l} (\mbox{Re}\,{\cal
E}+\Phi\overline{\Phi})\left(2{\cal E}_{uv}+ \displaystyle{\frac
{\alpha_u}\alpha}\,{\cal
E}_v+\displaystyle{\frac{\alpha_v}\alpha}\,{\cal
E}_u \right)-\left({\cal E}_u+2\overline{\Phi}\Phi_u\right) \,{\cal
E}_v-\left({\cal E}_v+2\overline{\Phi}\Phi_v\right)
\,{\cal E}_u=0\\[1ex]
(\mbox{Re}\,{\cal E}+\Phi\overline{\Phi})\left(2\Phi_{uv}+
\displaystyle{\frac{\alpha_u}\alpha}\Phi_v+
\displaystyle{\frac{\alpha_v}\alpha}\Phi_u\right)-\left({\cal E}_u+2\overline{\Phi}\Phi_u\right)\Phi_v-
\left({\cal E}_v+2\overline{\Phi}\Phi_v\right)\Phi_u=0\\[1ex]
\,\,\alpha_{uv}=0
\end{array}\right.
\end{equation}
Given a solution $(\alpha,\,{\cal E},\,\Phi)$ of (\ref{ErnstEqs}), the functions $H$, $\Omega$ and $A_y$, $A_z$ can be determined from  (\ref{EFPotentials}), while the function $f$ is determined by the relations
     \begin{equation}
    \label{CFactor}
\left\{\begin{array}{l}
\displaystyle{\frac {f_u}f}=\displaystyle{\frac{\alpha_{uu}}
{\alpha_u}}-\displaystyle{\frac {H_u}H}+
\displaystyle{\frac \alpha {2\alpha_u}}\left[\displaystyle{\frac
{\left\vert{\cal
E}_u+2\overline{\Phi}\Phi_u\right\vert^2}
{H^2}}+\displaystyle{\frac 4 H} \vert\Phi_u\vert^2\right]\\[3ex]
\displaystyle{\frac {f_v} f}=\displaystyle{\frac {\alpha_{vv}}
{\alpha_v}}-\displaystyle{\frac {H_v} H}+\displaystyle{\frac \alpha
{2\alpha_v}}
\left[\displaystyle{\frac {\left\vert{\cal
E}_v+2\overline{\Phi}\Phi_v\right\vert^2}
{H^2}}+\displaystyle{\frac 4 H}\vert\Phi_v\vert^2\right]
     \end{array}\right.
     \end{equation}

In the presence of electromagnetic fields we have no any analogues of  vacuum to vacuum Kramer-Neugebauer transformation and therefore, for construction of solutions for travelling gravitational and electromagnetic waves in some Kasner backgrounds we need to find some other transformations or ansatzes.

\section*{Travelling electromagnetic waves propagating in Kasner space-time}
\subsection*{On the Bonnor transformation}
The simplest way to construct the solutions for travelling electromagnetic waves could be an application to solutions for travelling gravitational waves discussed above of Bonnor transformation \cite{Bonnor:1961,SKMHH:2003} which maps (in hyperbolic case) a class of vacuum metrics admitting the Abelian two-dimensional orthogonally transitive isometry group with space-like orbits onto a class of electrovacuum solutions with the same symmetry but with diagonal metrics and electromagnetic potential which arise from a none-diagonal part of the original vacuum metric.

The applications of the Bonnor transformation to the solutions (\ref{VacuumLeftWaves}) and (\ref{VacuumRightWaves}) lead to the classes of solutions for waves with linear polarizations which depend on only one arbitrary real function determining the profile of this electromagnetic wave. It is interesting to note here that this Bonnor transformation  changes also the parameters of the background Kasner solution so that from the waves (\ref{VacuumLeftWaves}) and (\ref{VacuumRightWaves}) on the Kasner background with the exponents $(p_x,p_y,p_z)=(-\frac{3}{13},\frac{4}{13},\frac{12}{13})$
we obtain pure electromagnetic waves propagating through the symmetric Kasner background with the exponents $(p_x,p_y,p_z)=(-\frac{1}{3},\frac{2}{3},\frac{2}{3})$ along its axis of symmetry.

However, we do not follow this way, but instead we use some ansatz (obviously inspired by applicqation of Bonnor transformation to (\ref{VacuumLeftWaves}) or (\ref{VacuumRightWaves})) which leads to a class of solutions depending on \emph{two} arbitrary functions which determine the amplitudes of electromagnetic  waves of both polarizations.

\subsection*{Electromagnetic waves propagating through a symmetric Kasner background}
Let us consider a class of electrovacuum fields for which the metric on the orbits of the isometry group possess a plane symmetry and the non-zero components of electromagnetic potential depend only on the null coordinate $u=t-x$:
\begin{equation}\label{ansatzplus}
H=\alpha(u,v),\qquad \Omega=0,\qquad A_y=A_y(u),\qquad A_z=A_z(u).
     \end{equation}
This means that for the Ernst potentials for this class of fields we have
\begin{equation}
\fl\quad\left\{\begin{array}{l}
\mbox{Re}\,{\cal E}=-\alpha(u,v)-\phi_{\scriptscriptstyle{+}}(u)\overline{\phi}{}_{\scriptscriptstyle{+}}(u),\\[1ex]
\partial_u\mbox{Im}\,{\cal E}= i\left[\overline{\phi}{}_{\scriptscriptstyle{+}}(u)\, \phi_{\scriptscriptstyle{+}}\!{}^\prime(u)-
\phi_{\scriptscriptstyle{+}}(u)\overline{\phi}{}_{\scriptscriptstyle{+}}\!{}^\prime(u)
\right],
\end{array}\right.
\quad \Phi=\phi_{\scriptscriptstyle{+}}(u)\equiv A_y(u)-i A_z(u).
     \end{equation}
Substitution of these expressions for the Ernst potentials into the Ernst equations (\ref{ErnstEqs}) shows that these equations are satisfied for arbitrary complex function $\phi_{\scriptscriptstyle{+}}(u)$ (or, equivalently, for arbitrary real functions $A_y(u)$ and $A_z(u)$. Then the conformal factor $f$ can be calculated from (\ref{CFactor}) and the full metric takes the form
\begin{equation}\label{EVLeftWaves}
\qquad ds^2=-\frac{1}{\sqrt{\alpha}}\exp\left[2\int\limits_{u_0}^u \frac{\vert\phi_{\scriptscriptstyle{+}}\!{}^\prime(u)\vert^2
}{\alpha_u} du\right] \alpha_u \alpha_v du dv+\alpha(dy^2+dz^2)
\end{equation}
where the prime denotes a derivative. If the electromagnetic field depends only on the other null coordinate $v=t+x$, i.e. in the case $\Phi=\phi_{\scriptscriptstyle{-}}(v)$  we use the ansatz
\begin{equation}\label{ansatzminus}
H=\alpha(u,v),\qquad \Omega=0,\qquad A_y=A_y(v),\qquad A_z=A_z(v).
     \end{equation}
This leads to another class of solutions similar to (\ref{EVLeftWaves}):
\begin{equation}\label{EVRightWaves}
\qquad ds^2=-\frac{1}{\sqrt{\alpha}}\exp\left[2\int\limits_{v_0}^v \frac{\vert\phi_{\scriptscriptstyle{-}}\!{}^\prime(v)\vert^2}{\alpha_v}
dv\right] \alpha_u \alpha_v du dv+\alpha(dy^2+dz^2)
\end{equation}
where $\Phi=\phi_{\scriptscriptstyle{-}}(v)\equiv A_y(v)+i A_z(v)$ is an arbitrary complex function of $v$. Similarly to vacuum solutions (\ref{VacuumLeftWaves}) and (\ref{VacuumRightWaves}), the function $\alpha(u,v)$ (as the solution of the last equation in (\ref{ErnstEqs})) can be chosen in (\ref{EVLeftWaves}) and (\ref{EVRightWaves}) so that the surfaces $\alpha = const$ can be spase-like or time-like and therefore, we can choose for simplicity $\alpha=t$ or $\alpha=x$ respectively.

In the case $\alpha=t$, in the absence of electromagnetic fields, i.e. for $\phi_{\scriptscriptstyle{+}}(u)=0$ and $\phi_{\scriptscriptstyle{-}}(v)=0$, each of the solutions (\ref{EVLeftWaves}) and (\ref{EVRightWaves}) coincides with the symmetric Kasner solution and therefore, if $\phi_{\scriptscriptstyle{+}}(u)\ne 0$ and $\phi_{\scriptscriptstyle{-}}(v)\ne 0$, these solutions describe travelling electromagnetic waves propagating through (and interacting with) the symmetric Kasner background along its axis of symmetry ($x$-axis) in its right and left directions respectively. It is interesting to note that in the metrics  (\ref{EVLeftWaves}) and (\ref{EVRightWaves}) the electromagnetic field does not affect at all the part of metric which is transverse to the direction of waves propagation and therefore these are pure electromagnetic waves without any comoving gravitational wave component.

For the case of time-like $\alpha = const$, the ``twins'' of solutions (\ref{EVLeftWaves}) and (\ref{EVRightWaves}) representing some cylindrical waves with $\alpha = \rho$  propagating in some static axisymmetric background
have been found in \cite{Misra-Radhakrishna:1962} and mentioned also in the eq. (22.59) in the ``Cylindrical Waves'' chapter of \cite{SKMHH:2003}. However, it is necessary to note here, that the cylindrical wave interpretation of these solutions is not completely appropriate because of singular behaviour of these metrics on the axis $\rho=0$.

In our case with $\alpha=t$, from (\ref{EVLeftWaves}) and (\ref{EVRightWaves}) we obtain the solutions for plane-fronted electromagnetic waves propagating on the expandig (Kasner) background and the existence of Kasner initial singularity in these solutions is obviously inevitable from physical point of view. These solutions   can be useful in farther investigation of propagation and nonlinear interaction of travelling electromagnetic waves with a curved background as well as of collision  of these waves with arbitrary amplitudes with gravitational and electromagnetic solitons as well as with each other and with gravitational waves of arbitrary amplitudes. Besides that, the purpose of consideration of these waves in the context of the present paper is that a very simple ansatz (\ref{ansatzplus}) or (\ref{ansatzminus}), which leads respectively to the solutions (\ref{EVLeftWaves}) and (\ref{EVRightWaves}), can be generalized appropriately and used for construction of travelling massless bosonic waves through some (Kasner-like) cosmological backgrounds in some other (symmetry reduced) gravity models in four and higher dimensions.

\section*{Matrix hyperbolic Ernst equations for bosonic fields in string gravity}

\subsection*{Massless bosonic dynamics in string gravity}
We consider now the string gravity models in space-times with $D\ge 4$ dimensions for which the dynamics of massless bosonic fields is determined by the effective action
\begin{equation}
\label{StringFrame}
\fl{\cal S}=\!\!\displaystyle\int \! e^{-\widehat{\Phi}}\!\left\{\widehat{R}{}^{(D)}+\nabla_M \widehat{\Phi} \nabla^M\widehat{\Phi}
-\displaystyle\frac 1{12} H_{MNP} H^{MNP}
-\frac 12\sum\limits_{\mathfrak{p}=1}^n\displaystyle F_{MN}{}^{(\mathfrak{p})} F^{MN\,(\mathfrak{p})}\right\}\sqrt{- \widehat{G}}\,d^{D}x
\end{equation}
where $M,N,\ldots=1,2,\ldots,D$ and $\mathfrak{p}=1,\ldots n$; $\widehat{G}{}_{MN}$ possesses the ``most positive'' Lorentz signature. The string frame metric $\widehat{G}{}_{MN}$ and dilaton field $\widehat{\Phi}$ are related to the metric  $G_{MN}$ and dilaton  $\Phi$ in the Einstein frame as
\begin{equation}
\label{EinsteinFrame}
\widehat{G}{}_{MN}=e^{2\Phi} G_{MN},\qquad
\widehat{\Phi}=(D-2)\Phi.
\end{equation}
The components of a three-form $H$ and two-forms $F{}^{(\mathfrak{p})}$ are determined in terms of antisymmetric tensor field $B_{MN}$ and Abelian gauge field potentials $A_M{}^{(\mathfrak{p})}$ as
\[\begin{array}{l}
H_{MNP}=3\bigl(\partial_{[M} B{}_{NP]}-\sum\limits_{\mathfrak{p}=1}^n A{}_{[M}{}^{(\mathfrak{p})} F_{NP]}{}^{(\mathfrak{p})}\bigr),\\[0.5ex]
F_{MN}{}^{(\mathfrak{p})}=2\,\partial_{[M} A{}_{N]}{}^{(\mathfrak{p})},\qquad B_{MN}=-B_{NM}.
\end{array}
\]

\subsection*{Space-time symmetry ansatz}
The integrable hyperbolic reductions of the dynamical equations of the model (\ref{StringFrame}) arise from the assumption that the space-time admits $d=D-2$ commuting space-like Killing vector fields. All field components and potentials are assumed to be functions of time $t$ and one spatial coordinate $x$ (or, equivalently, of the null cone coordinates $u=t-x$ and $v=t+x$). It is assumed also that the metric components possess the structure
\begin{equation}\label{SMetric}
G_{MN}=\left(\begin{array}{ll}g_{\mu\nu}&0\\
0 & G_{ab}
\end{array}\right)\qquad
\begin{array}{l}
\mu,\nu,\ldots=1,2\\
a,b,\ldots=3,4,\ldots D
\end{array}
\end{equation}
while the components of gauge field potentials take the forms
\begin{equation}\label{BAfields}
B_{MN}=\left(\begin{array}{ll} 0& 0\\
0 & B_{ab}
\end{array}\right),\qquad
A_M{}^{(\mathfrak{p})}=\left(\begin{array}{l}0\\ A_a{}^{(\mathfrak{p})}
\end{array}\right).
\end{equation}
The coordinates $x^1,x^2$ can be chosen so that the metric $g_{\mu\nu}$ takes a conformally flat form $g_{\mu\nu} dx^\mu dx^\nu=f \, (-dt^2+dx^2)=-f\, du dv$ where  the conformal factor $f(u,v)> 0$.
The other metric and gauge field components can be combined into  three matrix functions
\begin{equation}\label{MVariables}
\mathcal{G}=e^{2\Phi}\Vert G_{ab}\Vert,\qquad
\mathcal{B}=\Vert B_{ab}\Vert,\qquad
\mathcal{A}=\Vert A_a{}^{(\mathfrak{p})}\Vert
\end{equation}
where $\mathcal{G}$ is a symmetric $d\times d$-matrix of string frame metric components, $\mathcal{B}$ is the antisymmetric $d\times d$-matrix of the components of gauge potential for the 3-form $H$
and $\mathcal{A}$ is a rectangular $d\times n$-matrix of components of $n$ Abelian vector gauge potentials.

\subsection*{Matrix Ernst potentials and hyperbolic matrix Ernst equations}
The space-time symmetry ansatz described above reduces the dynamical equations for the model (\ref{StringFrame}) to a completely integrable hyperbolic system which can be presented in different convenient forms.\footnote{In particular, in \cite{Alekseev:2009, Alekseev:2013}, a special $(2 d+n)\times (2 d+n)$-matrix form of these dynamical equations was found and used for a proof and exploration of integrable structure of these equations.} One of the forms of these equations represents a matrix analogue of the hyperbolic Ernst equations (see \cite{Alekseev:2009, Alekseev:2013}). In this form, any solution is determined by the $d\times d$-matrix of the Ernst-like potential $\mathcal{E}$, the rectangular $d\times n$-matrix $\mathcal{A}$ and a scalar function $\alpha$.

The matrix Ernst potential and the function $\alpha$  are defined in terms of matrix variables (\ref{MVariables}) and dilaton $\widehat{\Phi}$ by the following relations
\begin{equation}\label{MatrVar}
{\cal E}=\mathcal{G}+\mathcal{B}+{\cal A} {\cal A}^T,\qquad
\det\mathcal{G}\equiv e^{2\widehat{\Phi}}\alpha^2,\qquad \partial_u\partial_v\alpha=0,
\end{equation}
where the superscript ${}^T$ means a matrix transposition, the second equation in (\ref{MatrVar}) is a definition of the function $\alpha(u,v)>0$ and the last linear equation for $\alpha$ follows immediately from the symmetry reduced field equations for (\ref{StringFrame}).
In terms of the matrix potentials $\mathcal{E}$ and $\mathcal{A}$ and the function $\alpha$ the dynamical equations take the form
\begin{equation}\label{MErnstEqs}
\fl\hskip0ex\left\{\hskip-1ex\begin{array}{l}
2{\cal E}_{uv}+ \displaystyle{\frac
{\alpha_u}\alpha}\,{\cal
E}_v+\displaystyle{\frac{\alpha_v}\alpha}\,{\cal
E}_u -\left({\cal E}_u-2 \mathcal{A}_u \mathcal{A}{}^T\right)\mathcal{G}^{-1} \,{\cal
E}_v-\left({\cal E}_v-2 \mathcal{A}_v \mathcal{A}{}^T\right) \mathcal{G}^{-1}\,{\cal
E}_u=0\\[1ex]
2{\cal A}_{uv}+ \displaystyle{\frac
{\alpha_u}\alpha}\,{\cal
A}_v+\displaystyle{\frac{\alpha_v}\alpha}\,{\cal
A}_u -\left({\cal E}_u-2 \mathcal{A}_u \mathcal{A}{}^T\right)\mathcal{G}^{-1} \,{\cal
A}_v-\left({\cal E}_v-2 \mathcal{A}_v \mathcal{A}{}^T\right) \mathcal{G}^{-1}\,{\cal
A}_u=0\\[1ex]
\,\,\alpha_{uv}=0
\end{array}\right.
\end{equation}
where the suffices mean the derivatives and the matrix $\mathcal{G}=\frac12
(\mathcal{E}+\mathcal{E}^T)-\mathcal{A} \mathcal{A}^T$.

Given a solution $(\alpha,\,\mathcal{E},\,\mathcal{A})$ of (\ref{MErnstEqs}), the matrix functions $\mathcal{G}$ and $\mathcal{B}$ can be determined respectively from the symmetric and antisymmetric parts of matrix Ernst potential, while the string frame conformal factor $\widehat{f}$ is determined from the relations
\begin{equation}
    \label{CSFactor}
\fl\left\{\begin{array}{l}
\partial_u\log\left(\displaystyle{\frac{\alpha \widehat{f}}{\alpha_u\alpha_v}}\right) = \displaystyle{\frac{\alpha}{\alpha_u}}
\hbox{tr}\left[\displaystyle{\frac14}
({\cal E}_u-2{\cal A}_u{\cal A}^T)\, \mathcal{G}{}^{-1}({\cal E}_u^T-2 {\mathcal A}{\mathcal A}_u^T)\mathcal{G}{}^{-1}+
\mathcal{A}_u^T \mathcal{G}^{-1}\mathcal{A}_u \right]
\\[3ex]
\partial_v\log\left(\displaystyle{\frac{\alpha \widehat{f}}{\alpha_u\alpha_v}}\right) =\displaystyle{\frac{\alpha}{\alpha_v}}
\hbox{tr}\left[\displaystyle{\frac14}
({\cal E}_v-2{\cal A}_v{\cal A}^T)\, \mathcal{G}{}^{-1}({\cal E}_v^T-2 {\mathcal A}{\mathcal A}_v^T)\mathcal{G}{}^{-1}+
\mathcal{A}_v^T \mathcal{G}^{-1}\mathcal{A}_v \right]
     \end{array}\right.
     \end{equation}
and the conformal factor in the Einstein frame is $f=e^{-2\Phi} \widehat{f}$.

It is useful to note that for any solution $(\alpha,\,\mathcal{E},\,\mathcal{A})$ of of matrix Ernst equations (\ref{MErnstEqs}) and corresponding $\mathcal{G}$ and $\mathcal{B}$, there exist two dual matrix potentials $\widetilde{\mathcal{B}}$ and $\widetilde{\mathcal{A}}$ such that
\begin{equation}\label{SDuals}
\fl\left\{\begin{array}{l}
\widetilde{\mathcal{B}}{}_u= -\alpha \mathcal{G}^{-1}(\mathcal{B}{}_u-{\cal A}{}_u {\cal A}^T+{\cal A} {\cal A}^T_u)\mathcal{G}^{-1},\\[1ex]
\widetilde{\mathcal{B}}{}_v= \alpha \mathcal{G}^{-1}(\mathcal{B}{}_v-{\cal A}{}_v {\cal A}^T+{\cal A} {\cal A}^T_v)\mathcal{G}^{-1},
\end{array}\right.\qquad
\left\{\begin{array}{l}
\widetilde{\mathcal{A}}{}_u=-\alpha \mathcal{G}^{-1} \mathcal{A}{}_u+\widetilde{\mathcal{B}}\,\mathcal{A}{}_u,\\[1ex]
\widetilde{\mathcal{A}}{}_v=\alpha \mathcal{G}^{-1} \mathcal{A}{}_v+\widetilde{\mathcal{B}}\,\mathcal{A}{}_v,
\end{array}\right.
\end{equation}
Just these dual potentials will be used in our construction of the alternative Ernst potential and matrix generalization of the Kramer-Neugebauer transformation.

\section*{Hyperbolic ``vacuum'' matrix  Ernst equation and Kramer-Neugebauer transformation of its solution space}

\subsection*{Matrix analogue of vacuum Ernst equation}
From mathematical point of view, the matrix analogue of pure vacuum Ernst equation of General Relativity arises from matrix Ernst equations (\ref{MErnstEqs}) in the corresponding ``vacuum'' case for which $\mathcal{A}=0$, i.e.if  all Abelian vector gauge fields vanish.  In this ``vacuum'' case, however, the set of dynamical variables consist of the string frame metric $\mathcal{G}$ and the antisymmetric matrix potential $\mathcal{B}$ which can be combined (as in general case (\ref{MatrVar}), into a matrix Ernst potential and the dynamical equations take the form
\begin{equation}\label{MErnstEqn}
\fl\hskip4ex\left.\left\{\hskip-1ex\begin{array}{l}
2{\cal E}_{uv}+ \displaystyle{\frac
{\alpha_u}\alpha}\,{\cal
E}_v+\displaystyle{\frac{\alpha_v}\alpha}\,{\cal
E}_u -{\cal E}_u\mathcal{G}^{-1} \,{\cal
E}_v-{\cal E}_v\mathcal{G}^{-1}\,{\cal
E}_u=0,\\[1ex]
\,\,\alpha_{uv}=0.
\end{array}\right.\quad\right\Vert\quad
\begin{array}{l}
{\cal E}=\mathcal{G}+\mathcal{B}\\[1ex]
\det\mathcal{G}\equiv e^{2\widehat{\Phi}}\alpha^2
\end{array}
\end{equation}
The defined above dual matrix potential $\widetilde{\mathcal{B}}$ is   determined in terms of these variables as
\begin{equation}\label{SVDuals}
\widetilde{\mathcal{B}}{}_u= -\alpha \mathcal{G}^{-1}\mathcal{B}{}_u\mathcal{G}^{-1},\qquad
\widetilde{\mathcal{B}}{}_v= \alpha \mathcal{G}^{-1}\mathcal{B}{}_v\mathcal{G}^{-1}.
\end{equation}

\subsection*{Alternative vacuum Ernst potential and  Kramer-Neugebauer transformation}
As it can be checked by a direct calculation, for any solution $(\alpha,\mathcal{E})$ of the matrix Ernst equation (\ref{MErnstEqn}) the corresponding matrices $\mathcal{G}$ and $\mathcal{B}$ the following alternative combination of the corresponding matrices $\mathcal{G}$ and $\mathcal{B}$ also should satisfy (\ref{MErnstEqn}):
\begin{equation}\label{AltMEPotential}
{\cal E}=\alpha\, \mathcal{G}^{-1}+\widetilde{B}
\end{equation}
Therefore, given a solution $(\alpha,{\cal E})$ of (\ref{MErnstEqn}), we can construct two ``vacuum'' solutions corresponding to different ``interpretations'' (\ref{MErnstEqn}) and (\ref{AltMEPotential}) of the matrix Ernst potential:
\begin{equation}\label{VSTransform}
\fl\left\{\begin{array}{l}
\mathcal{G}_{(1)}=\displaystyle{\frac12}(\mathcal{E}+\mathcal{E}^T),\\
\mathcal{B}_{(1)}=\displaystyle{\frac12}(\mathcal{E}-\mathcal{E}^T),
\end{array}\right.\qquad
\left\{\begin{array}{lcl}
\mathcal{G}_{(2)}=\alpha\, \mathcal{G}_{(1)}^{-1},&&
\widehat{f}{}_{(2)}= \alpha^{\frac {d-4}{4}}e^{-\widehat{\Phi}{}_{(1)}} \widehat{f}{}_{(1)}\\[1.75ex]
d\mathcal{B}_{(2)}=-\alpha\, \mathcal{G}^{-1}_{(1)}\, {}^{\ast}\!d\mathcal{B}_{(1)}\,\, \mathcal{G}^{-1}_{(1)},&&
e^{\widehat{\Phi}{}_{(2)}}=\alpha^{\frac {d-4}{2}}e^{-\widehat{\Phi}{}_{(1)}}
\end{array}\right.
\end{equation}
Thus, we have the transformation acting on the space of solutions of the described above symmetry reduced ``vacuum'' ($\mathcal{A}=0$) string gravity model (\ref{StringFrame})  which takes the form
\begin{equation}\label{SNKTransform}
\{\mathcal{G}_{(1)},\,\mathcal{B}_{(1)},\Phi_{(1)}\}\quad\longleftrightarrow\quad \{\mathcal{G}_{(2)},\,\mathcal{B}_{(2)},\Phi_{(2)}\}
\end{equation}
This transformation generalizes to the case of ``vacuum'' string gravity model and, more precisely, to the case of its hyperbolic symmetry reductions, the transformations discovered by Kramer and Neugebauer for stationary axisymmetric  vacuum solutions in General Relativity (see \cite{SKMHH:2003} for more details and references).

\section*{Travelling ``vacuum'' massless bosonic waves which are fixed points\\ of Kramer-Neugebauer transformation}
To construct the solutions  for the travelling ``vacuum'' ($\mathcal{A}=0$) bosonic waves, we use the same way as it was used in one of the previous sections of this paper for construction of pure vacuum travelling gravitational waves in General Relativity -- the solutions (\ref{VacuumLeftWaves}) and (\ref{VacuumRightWaves}). Namely, we consider the solutions of matrix Ernst equations (\ref{MErnstEqn}) which are fixed points of the transformation (\ref{SNKTransform}). These solutions should satisfy
\begin{equation}\label{SInvariance}
\mathcal{G}=\alpha \mathcal{G}^{-1},\quad
\widetilde{\mathcal{B}}=\mathcal{B}.
\end{equation}
The first of these equations, together with the condition of positive signature of the matrix $\mathcal{G}$, means that this matrix is proportional to a unit matrix and thus, we have
\[\mathcal{G}=\sqrt{\alpha}\, I\quad \Rightarrow\quad \det \mathcal{G}=\alpha^{d/2}\quad\Rightarrow\quad
e^{\widehat{\Phi}}=\alpha^{\frac{d-4}4}
\]
Then, in accordance with the definition (\ref{SVDuals}) of $\widetilde{B}$, the second of the invariance conditions (\ref{SInvariance}) means that $\mathcal{B}=\mathcal{B}_{\scriptscriptstyle{-}}(v)$ where $\mathcal{B}_{\scriptscriptstyle{-}}(v)$ is an arbitrary antisymmetric $d\times d$-matrix function of $v=t+x$. A direct substitution of this solution into the matrix Ernst equation (\ref{MErnstEqn}) shows that this equation is satisfied. Calculating the conformal factor from (\ref{CSFactor}) we obtain the solution in the string frame in the form
\begin{equation}\label{SVRightWaves}
\fl \quad
\begin{array}{l}
ds_D^2=-\alpha^{\frac{d}{16}-1}\exp\left[-\displaystyle \int\limits_{v_0}^v \displaystyle\frac{\hbox{tr} \bigl(\mathcal{B}_{\scriptscriptstyle{-}}^\prime\! \cdot \mathcal{B}_{\scriptscriptstyle{-}}^\prime\!\bigr)}{4\alpha_v} dv\right] \alpha_u  \alpha_v du dv+\sqrt{\alpha}\Bigl[{dx^3}^2+\ldots +{dx^{D}}^2\Bigr],\\[4ex]
\mathcal{A}=0,\quad\mathcal{B}=\mathcal{B}_{\scriptscriptstyle{-}}\!(v),\quad \widehat{\Phi}=\frac{d-4}4\,\log \alpha,\quad \alpha=\alpha(u,v),\quad \alpha_{uv}=0
\end{array}
\end{equation}
where primes denote the derivatives. Besides that, using instead of (\ref{SInvariance}), a different invariance condition $\mathcal{G}=\alpha \mathcal{G}^{-1}$ and $\widetilde{\mathcal{B}}=-\mathcal{B}$, we obtain another
class of solutions
\begin{equation}
\label{SVLeftWaves}
\fl \quad\begin{array}{l}
ds_D^2=-\alpha^{\frac{d}{16}-1}\exp\left[-\displaystyle \int\limits_{u_0}^u \displaystyle\frac{\hbox{tr} \bigl(\mathcal{B}_{\scriptscriptstyle{+}}^\prime\! \cdot \mathcal{B}_{\scriptscriptstyle{+}}^\prime\!\bigr)}{4\alpha_u} du\right] \alpha_u  \alpha_v du dv+\sqrt{\alpha}\Bigl[{dx^3}^2+\ldots +{dx^{D}}^2\Bigr],\\[3ex]
\mathcal{A}=0,\quad\mathcal{B}=\mathcal{B}_{\scriptscriptstyle{+}}\!(u),\quad \widehat{\Phi}=\frac{d-4}4\,\log \alpha,\quad \alpha=\alpha(u,v),\quad \alpha_{uv}=0.
\end{array}
\end{equation}

Similarly to the solutions constructed in previous sections, for different choices of  $\alpha$ as a solution $\alpha_{uv}=0$ for which  $\alpha=const$ are time-like or space-like surface, the transformations  $u\to h(u)$, $v\to g(v)$ allow to choose $t=\alpha$ or $x=\alpha$ respectively.

It is easy to see that in the case $\alpha=t$, for $\mathcal{B}_{\scriptscriptstyle{-}}\!(v)=0$ and $\mathcal{B}_{\scriptscriptstyle{+}}\!(u)=0$ the solutions (\ref{SVRightWaves}) and (\ref{SVLeftWaves}) respectively reduce to a $D$-dimensional Kasner-like background and therefore, for $\mathcal{B}_{\scriptscriptstyle{-}}\!(v)\ne 0$ and $\mathcal{B}_{\scriptscriptstyle{+}}\!(u)\ne 0$ these solutions describe
some travelling waves of a three-form field $H$ from the bosonic sector of string gravity model (\ref{StringFrame}). The arbitrary antisymmetric matrix functions
$\mathcal{B}_{\scriptscriptstyle{-}}\!(v)$ and $\mathcal{B}_{\scriptscriptstyle{+}}\!(u)$ describe arbitrary amplitudes of these waves and arbitrary states of their polarizations. It seems interesting to note also that in these travelling wave solutions the part of metric transversal to the direction of waves propagation ($x$-axis) remains unperturbed and therefore, these waves do not comprise pure gravitational and dilaton wave components.

\section*{Travelling waves of bosonic gauge fields in the Kasner-like background}
To construct the solutions of (\ref{MErnstEqs}) which describe the travelling waves for vector gauge fields, we follow the same way which we used in previous sections for construction of pure electromagnetic waves propagating on the symmetric Kasner background in the Einstein-Maxwell theory. Namely, we assume that in these waves the part of metric transversal to the direction of wave propagation is not perturbed, i.e. the gravitational component of these waves is absent. As before, we assume also for the string frame spatial metric a Kasner-like form with specially chosen (equal to each other) exponents and restrict our consideration by the case in which the dual potential $\widetilde{\mathcal{B}}$ vanishes:
\begin{equation}
\label{S2Ansatz}
\qquad\mathcal{G}=\alpha\,I,\qquad \widetilde{\mathcal{B}}=0.
\end{equation}
With these two assumptions we obtain immediately from (\ref{SDuals}) the relations:
\begin{equation}
\label{SRDuals}
\left\{\begin{array}{l}
\mathcal{B}{}_u={\cal A}{}_u {\cal A}^T-{\cal A} {\cal A}^T_u,\\[1ex]
\mathcal{B}{}_v={\cal A}{}_v {\cal A}^T-{\cal A} {\cal A}^T_v,
\end{array}\right.\qquad
\left\{\begin{array}{l}
\widetilde{\mathcal{A}}{}_u=-\mathcal{A}{}_u,\\[1ex]
\widetilde{\mathcal{A}}{}_v=\mathcal{A}{}_v.
\end{array}\right.
\end{equation}
The last two relations imply the condition $\mathcal{A}{}_{uv}=0$, but for travelling waves we should restrict ourselves to one of the cases $\mathcal{A}{}_{v}=0$ or $\mathcal{A}{}_{u}=0$ which imply respectively the conditions $\mathcal{B}{}_{v}=0$ or $\mathcal{B}{}_{u}=0$. In the first case, we obtain
\[\fl\qquad\mathcal{A}=\mathcal{A}_{\scriptscriptstyle{+}}\!(u),\qquad
\mathcal{B}=\mathcal{B}_{\scriptscriptstyle{+}}\!(u),\qquad
\mathcal{B}_{\scriptscriptstyle{+}}^\prime\!(u)={\cal A}_{\scriptscriptstyle{+}}^\prime\!(u) {\cal A}_{\scriptscriptstyle{+}}^T\!(u)-{\cal A}_{\scriptscriptstyle{+}}\!(u) {\cal A}_{\scriptscriptstyle{+}}^\prime{}^T\!(u),
\]
while in the second case we have
\[\fl\qquad\mathcal{A}=\mathcal{A}_{\scriptscriptstyle{-}}\!(v),\qquad
\mathcal{B}=\mathcal{B}_{\scriptscriptstyle{-}}\!(v),\qquad
\mathcal{B}_{\scriptscriptstyle{v}}^\prime\!(v)={\cal A}_{\scriptscriptstyle{-}}^\prime\!(v) {\cal A}_{\scriptscriptstyle{-}}^T\!(v)-{\cal A}_{\scriptscriptstyle{-}}\!(v) {\cal A}_{\scriptscriptstyle{-}}^\prime{}^T\!(v).
\]
Hence, given the matrix $\mathcal{A}_{\scriptscriptstyle{+}}(u)$  (in the first case) or the matrix $\mathcal{A}_{\scriptscriptstyle{-}}(v)$ (in the second case), one can calculate from the above expressions the corresponding  three-form potential $\mathcal{B}_{\scriptscriptstyle{+}}\!(u)$ or $\mathcal{B}_{\scriptscriptstyle{-}}\!(v)$ respectively.
Direct substitution of all these expressions into the matrix Ernst equations (\ref{MErnstEqs}) shows that these equations are satisfied for arbitrary matrix functions $\mathcal{A}_{\scriptscriptstyle{+}}\!(u)$ or $\mathcal{A}_{\scriptscriptstyle{-}}\!(v)$.
After calculation of the conformal factors from the expressions (\ref{CSFactor}) we obtain a class of solutions with string frame metrics
\begin{equation}
\label{SLeftWaves}
\fl \quad\begin{array}{l}
ds_D^2=-\alpha^{\frac{d}{4}-1}\exp\left[\displaystyle \int\limits_{u_0}^u \displaystyle\frac{\hbox{tr} \bigl(\mathcal{A}_{\scriptscriptstyle{+}}^\prime{}^T\! \cdot \mathcal{A}_{\scriptscriptstyle{+}}^\prime\!\bigr)}{\alpha_u} du\right] \alpha_u  \alpha_v du dv+\alpha\Bigl[{dx^3}^2+\ldots +{dx^{D}}^2\Bigr],\\[3ex]
\mathcal{A}=\mathcal{A}_{\scriptscriptstyle{+}}(u),
\quad\mathcal{B}=\displaystyle\int\limits_{u_0}^u
\left[{\cal A}_{\scriptscriptstyle{+}}^\prime\!\cdot {\cal A}_{\scriptscriptstyle{+}}^T\!-{\cal A}_{\scriptscriptstyle{+}}\! \cdot{\cal A}_{\scriptscriptstyle{+}}^\prime{}^T\!\right] du,\quad
\widehat{\Phi}=\frac{d-2}2\,\log \alpha.
\end{array}
\end{equation}
where ${\cal A}_{\scriptscriptstyle{+}}(u)$ is an arbitrary $d\times n$-matrix function; $\alpha=\alpha(u,v)$ and $\alpha_{uv}=0$.
The other class of solutions is
\begin{equation}
\label{SRightWaves}
\fl \quad\begin{array}{l}
ds_D^2=-\alpha^{\frac{d}{4}-1}\exp\left[\displaystyle \int\limits_{v_0}^v \displaystyle\frac{\hbox{tr} \bigl(\mathcal{A}_{\scriptscriptstyle{-}}^\prime{}^T\! \cdot \mathcal{A}_{\scriptscriptstyle{-}}^\prime\!\bigr)}{\alpha_v} dv\right] \alpha_u  \alpha_v du dv+\alpha\Bigl[{dx^3}^2+\ldots +{dx^{D}}^2\Bigr],\\[3ex]
\mathcal{A}=\mathcal{A}_{\scriptscriptstyle{-}}(v),
\quad\mathcal{B}=\displaystyle\int\limits_{v_0}^v
\left[{\cal A}_{\scriptscriptstyle{-}}^\prime\!\cdot {\cal A}_{\scriptscriptstyle{-}}^T\!-{\cal A}_{\scriptscriptstyle{-}}\! \cdot{\cal A}_{\scriptscriptstyle{-}}^\prime{}^T\!\right] dv,\quad
\widehat{\Phi}=\frac{d-2}2\,\log \alpha.
\end{array}
\end{equation}
where ${\cal A}_{\scriptscriptstyle{-}}(v)$ is an arbitrary $d\times n$-matrix function; $\alpha=\alpha(u,v)$ and $\alpha_{uv}=0$.

Similarly to our comments of previous solutions, we note that for different choices of  $\alpha$ as a solution $\alpha_{uv}=0$ for which  $\alpha=const$ are time-like or space-like surface, the transformations  $u\to h(u)$, $v\to g(v)$ allow to choose $t=\alpha$ or $x=\alpha$ respectively. In the case $\alpha=t$, for $\mathcal{A}_{\scriptscriptstyle{+}}\!(u)=0$ and $\mathcal{A}_{\scriptscriptstyle{-}}\!(v)=0$ the solutions (\ref{SLeftWaves}) and (\ref{SRightWaves}) respectively reduce to a $D$-dimensional Kasner-like background and therefore, for $\mathcal{A}_{\scriptscriptstyle{+}}\!(u)\ne 0$ and $\mathcal{A}_{\scriptscriptstyle{-}}\!(v)\ne 0$ these solutions describe travelling waves of arbitrary number $n$ of Abelian vector gauge fields from the bosonic sector of string gravity (\ref{StringFrame}). The arbitrary $d\times n$-matrix functions
$\mathcal{A}_{\scriptscriptstyle{+}}\!(u)$ and $\mathcal{A}_{\scriptscriptstyle{-}}\!(v)$ describe arbitrary amplitudes of these waves (propagating respectively in the positive and negative directions of $x$-axis) and arbitrary states of their polarizations. For these waves the part of metric transversal to the direction of propagation ($x$-axis) remains unperturbed and therefore, these waves also do not comprise pure gravitational and dilaton wave components, but some ``induced'' wave components of the three-from field with the  corresponding potentials $\mathcal{B}_{\scriptscriptstyle{+}}\!(u)$ or $\mathcal{B}_{\scriptscriptstyle{-}}\!(v)$ (depending on the choice of $\mathcal{A}_{\scriptscriptstyle{+}}\!(u)$ and $\mathcal{A}_{\scriptscriptstyle{-}}\!(v)$ respectively) presents in these solutions.

\section*{Summary of results and conclusions}

In this paper, we constructed several classes of exact travelling wave solutions
(i.e. the solutions for waves which propagate in a given space-time region in certain direction without any scattering, caustics and singularities) for gravitational and electromagnetic waves in General Relativity as well as for massless bosonic fields in some string gravity models in four and higher dimensions. Each of these classes of solutions (different from pp-waves) depend on a set of arbitrary functions of a null coordinate which allow to consider the waves with arbitrary profiles and states of polarizations. The solutions of these classes  describe propagation of these travelling waves in some expanding spatially homogeneous Kasner-like space-times.

From mathematical point of view, it is interesting that these classes of solutions were obtained as fixed points of (generalized) Kramer-Neugebauer transformations of the solution spaces of vacuum and certain ``vacuum-like'' hyperbolic integrable symmetry reductions respectively of vacuum Einstein equations in General Relativity and of dynamical equations for massless bosonic fields in string gravity models in four and higher dimensions.
The ``source'' of the other travelling wave solutions found in this paper is the ansatz which restricts the consideration of solutions of the hyperbolic Ernst equations in General relativity and matrix Ernst equations in string gravity by the solutions for which the part of spatial metric transversal to the direction of wave propagation coincides with that for the background solution.

From physical point of view, it is worth to note that the integrability of the hyperbolic Ernst equations and generalized (matrix) hyperbolic Ernst equations which govern the nonlinear behaviour of the waves considered in this paper as well as very simple form of all solutions constructed here, provide us with some effective methods (based on the modern theory of integrable systems) for consideration of many physically interesting questions concerning various aspects of the nonlinear interactions of these waves propagating in curved space-times. In particular,
\begin{itemize}
\item{application to these solutions of various global symmetry transformations such as Harrison and Bonnor transformations allow to construct various solutions which also include some arbitrary functions of null variable and which describe travelling gravitation and electromagnetic waves propagating on curved cosmological backgrounds not only of Kasner's but also of some other types.}
\item{Applications to these solutions of vacuum \cite{Belinski-Zakharov:1978} and electrovacuum \cite{Alekseev:1980} soliton generating techniques allow to consider the interaction of these waves (with arbitrary profiles) respectively with gravitational or gravitational and electromanetic solitons on the same Kasner background.}
\item{A collision of such travelling waves (e.g. the traveling waves of the sandwich type) also represent certain physically interest. In particular, it seems interesting to analyze the gravitational waves which may be created as a result of collision of pure electromagnetic or other massless bosonic travelling waves colliding with arbitrary amplitudes on the expanding spatially homogeneous background. The general approach to solution of the corresponding characteristic initial value problems was developed in \cite{Alekseev:2001, Alekseev-Griffiths:2001, Alekseev-Griffiths:2004}.}
\end{itemize}

However, the questions of collision and nonlinear interaction of waves are not in the scope of this paper, and it is expected that these will be the subject of farther studies.

\section*{Acknowledgments}
This work was supported in part by the Russian Foundation for Basic Research
(Grants No. 14-01-00049 and No. 14-01-00860) and the program ``Fundamental
problems of Nonlinear Dynamics'' of the Russian Academy of Sciences.

\end{document}